%% file: chi-dashboard.tex
 \documentclass[sigconf,screen]{acmart} 

\input{commands}

\AtBeginDocument{%
  }

\copyrightyear{2026}
\acmYear{2026}
\setcopyright{cc}
\setcctype{by}
\acmConference[AVI '26]{Proceedings of the 2026 International Conference on Advanced Visual Interfaces}{June 08--12, 2026}{Venice, Italy}
\acmBooktitle{Proceedings of the 2026 International Conference on Advanced Visual Interfaces (AVI '26), June 08--12, 2026, Venice, Italy}
\acmDOI{10.1145/3811427.3811529}
\acmISBN{979-8-4007-2342-1/2026/06}

\begin{document}

\title{InvestChat: Exploring Multimodal Interaction via Natural Language, Touch, and Pen in an Investment Dashboard}

\author{Sarah Lykke Tost}
\orcid{0009-0009-4625-9564}
\affiliation{%
  \institution{Aarhus University}
  \city{Aarhus}
  \country{Denmark}
}

\author{Adson Lucas de Paiva Sales}
\orcid{0009-0009-8180-6586}
\affiliation{%
  \institution{Aarhus University}
  \city{Aarhus}
  \country{Denmark}
}

\author{Henrik Østergaard}
\orcid{0009-0002-9227-3569}
\affiliation{%
  \institution{Aarhus University}
  \city{Aarhus}
  \country{Denmark}
}

\author{Vaishali Dhanoa}
\orcid{0000-0002-0493-8616}
\affiliation{%
  \institution{Aarhus University}
  \city{Aarhus}
  \country{Denmark}
}
\affiliation{%
  \institution{TU Wien}
  \city{Vienna}
  \country{Austria}
}

\author{Gabriela Molina León}
\orcid{0000-0002-9223-2022}
\affiliation{%
  \institution{Aarhus University}
  \city{Aarhus}
  \country{Denmark}
}

\renewcommand{\shortauthors}{Tost et al.}

\begin{abstract}
\input{sections/00-abstract}
\end{abstract}


\begin{teaserfigure}
  \centering
  \subfigure[Main Area of InvestChat]{
  \includegraphics[width=0.34\textwidth]{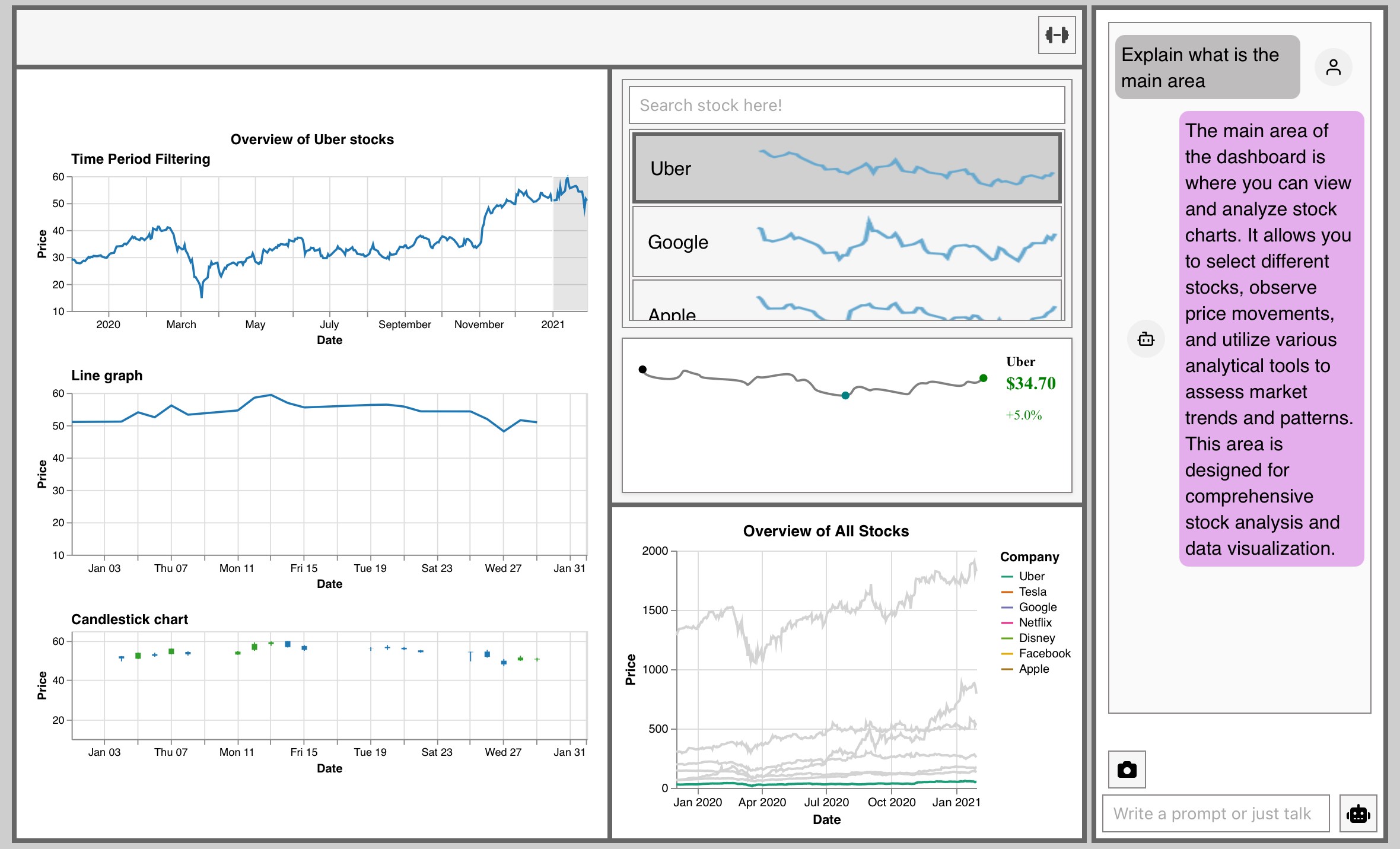}
  \label{fig:main}
  } 
  \subfigure[Training Area of InvestChat]{
  \includegraphics[width=0.34\textwidth]{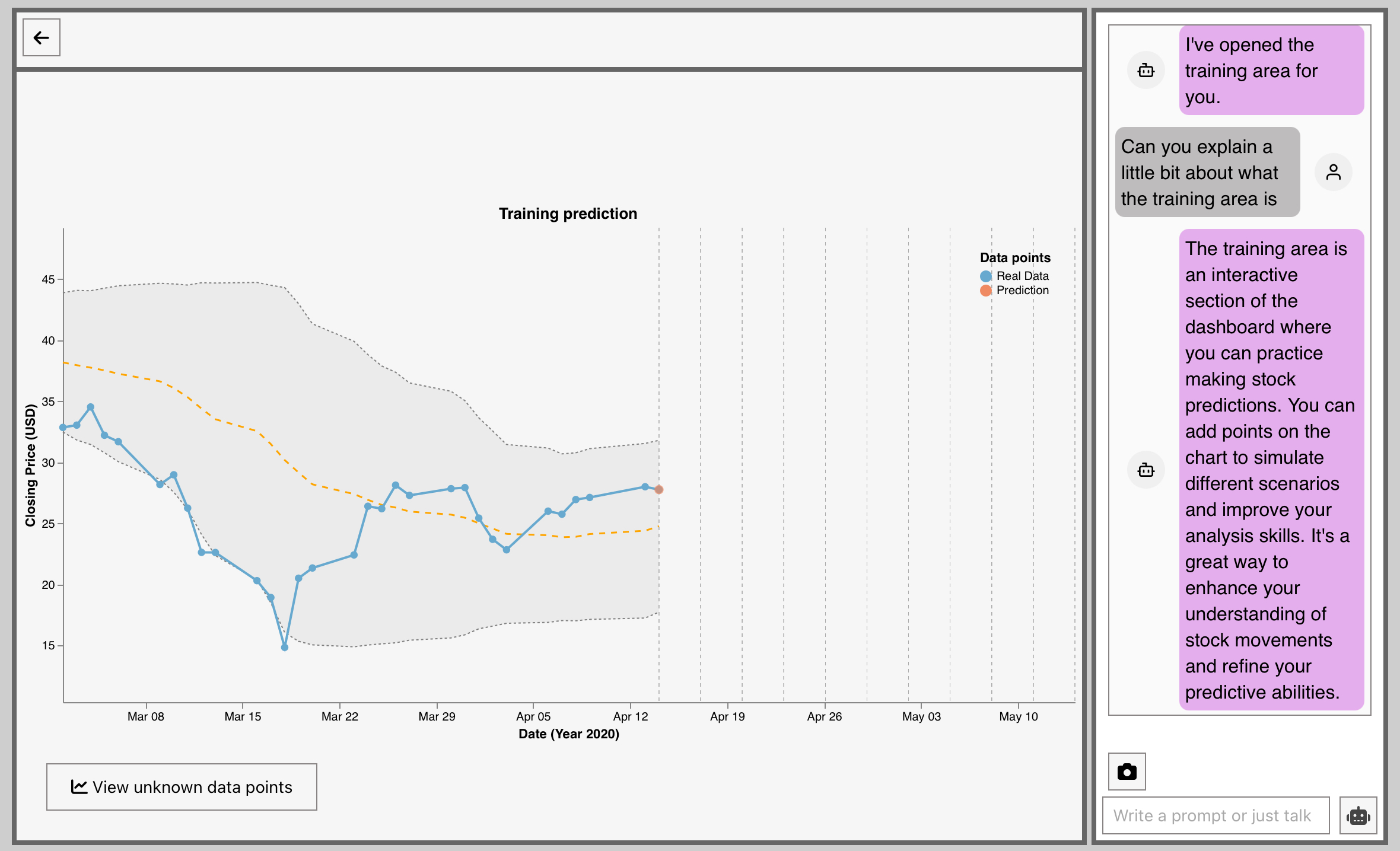}
  \label{fig:training}
  } 
  \caption{InvestChat Areas: (a)
  Stock overview with KPI and three different visualizations for the selected stock.
  (b) Prediction area to compare user predictions with historical data.
  Both areas include a navigation bar and a chat box for LLM interaction.
  }
  \Description{In the Main Area, the first two panels show multiple visualizations, while the third shows the chat history with two messages. The first is from a user and says ``Explain what is the main area.'' The second shows the system answer: ``The main area of the dashboard is where you can view and analyze stock charts. It allows you to select different stocks, observe price movements, and utilize various analytical tools to assess market trends and patterns. This area is designed for comprehensive stock analysis and data visualization. In the Training Area, we can interact through a chat with the LLM via natural language. We can give commands to the LLM, e.g., delete all added predicted points, navigate to Main Area, hide or show prediction points for comparison. As well as asking questions to form the prediction either through natural language or through the screenshot functionality, e.g., is the next trend upwards?''}
  \label{fig:minipage}
\end{teaserfigure}


\maketitle

\section{Introduction}
\input{sections/01-introduction}

\section{Related Work}
\input{sections/02-related-work}

\section{Designing InvestChat}
\input{sections/04-design}

\section{Evaluation}
\input{sections/06-evaluation}

\begin{acks}
We have used Generative AI as a tool for code refactoring and to provide a better structure to the implementation.
This work was partially supported by Villum Investigator grant VL-54492 by Villum Fonden.
Any opinions, findings, and conclusions expressed in this material are those of the authors and do not necessarily reflect the views of the funding agency.
\end{acks}

\bibliographystyle{ACM-Reference-Format}
\bibliography{references}


\end{document}

%% file: commands.tex
\usepackage{color} 

\usepackage{subfigure}
\usepackage{subcaption}
\usepackage{graphicx}

%% file: sections/00-abstract.tex
 
We designed and implemented \textit{InvestChat}, a multimodal tablet-based application that supports stock market exploration with multiple coordinated views and an LLM-powered chat. 
We evaluated the application with 12 novice investors. 
Our findings suggest that combining natural language, touch, and pen input during stock market exploration facilitates user engagement. Participants leveraged the modalities in complementary ways, enjoying the freedom of choice and finding natural language most effective.

%% file: sections/01-introduction.tex
Investment has long been a subject of interest and concern, particularly among youth \cite{PeopleInvestMore}. Currently, young investors use digital platforms on desktop computers or mobile devices. However, not everyone is tech-savvy, and investing 
requires careful consideration. Simultaneously, mobile applications often offer limited interaction capabilities, primarily through touch, whereas desktop computers are used mostly in static work settings. As investing often occurs during leisure time, tablets offer an ideal middle ground. Yet, few investment applications are designed for tablets, and most rely on mouse, keyboard, or touch input only \cite{Multimodal_Interaction_A_Review}.
To bridge this gap, we designed \textit{InvestChat}, a multimodal dashboard for investment data exploration and training on tablets.
\textit{InvestChat} provides users multiple ways to interact with investment data, namely through touch, stylus, and natural language (text \& speech). This multimodal approach aims at increasing freedom of expression \cite{InChorus} and reducing the gap between user intent and execution when interacting with data visualizations \cite{BeyondMouse}. Therefore, we designed \textit{InvestChat} to investigate: 
\textit{How can multimodal interaction facilitate engagement for novice investors during dashboard exploration on a tablet device?}

In an evaluation with 12 novice investors,
\textit{InvestChat} facilitated user engagement by supporting direct manipulation via touch and pen, as well as commands and querying via text and speech. Novices found \textit{InvestChat} highly usable, 
and often interacted with speech in open-ended tasks. 
We contribute
(1) the design and implementation of \textit{InvestChat}, 
and (2) insights from a study with novice investors. 

%% file: sections/02-related-work.tex
Over the past decade, InfoVis researchers have explored alternatives to mouse and keyboard interaction to reduce the gap between user intent and execution, to enable people to focus on analytical tasks \cite{BeyondMouse}. The alternatives are modalities considered  \emph{natural}, such as touch, pen, mid-air gestures, and speech, to simplify manipulation and lower cognitive load \cite{TouchPivot, BeyondMouse}.
Input modalities are often combined in multimodal solutions, as no single modality suits all contexts \cite{Multimodal-interaction, MMI-input} and different modalities offer complementary strengths \cite{Affordances}. For instance, touch gestures enable direct and expressive manipulation, even though they require proximity and have limited precision, whereas speech interaction supports high-level commands from a distance, but is less suitable for interactions associated with specific screen coordinates \cite{Talk_To_The_Wall}. Combining mouse and speech interaction increases the flow of visual analysis \cite{Chowdhury_dashboard_multimodal}. Yet, interaction designers need to choose modalities carefully to avoid increasing the cognitive load and development complexity \cite{Multimodal_Interaction_A_Review}.

%% file: sections/04-design.tex
We conducted a formative study with 49 university students, from Business \& IT, with experience or interest in investing, to understand current investment behaviors and challenges. When asked about what challenges they encountered when investing, people often reported difficulties related to a lack of information and domain knowledge. The study also showed that tablets were the least-used investment device, with only one person reporting usage. Additionally, 27 participants expressed no experience or interest in alternative modalities such as voice, touch, or pen input. Based on this feedback, we designed a multimodal dashboard to help novices build investing knowledge through an integrated LLM assistant and a dedicated training area. At the same time, we wished to encourage and challenge novices by employing a tablet-based platform and touch, stylus, and natural language (text \& speech) interactions, enabling new and more natural ways to explore investment data.
Our design choices were guided by relevant literature, especially Interchat \cite{InterChat}, which inspired LLM-driven chat interactions, and TimeFork \cite{TimeFork}, which informed time-series visualizations.

\emph{InvestChat} consists of two areas, as shown in \autoref{fig:minipage}.
The \textbf{Main Area} (Fig.~\autoref{fig:main}) focuses on stock visualization and exploration. 
Here, users can select stocks, filter time periods, and view multiple charts at once. 
This area also integrates an LLM-based chat box for guidance and data interaction.
The \textbf{Training Area} (Fig.~\autoref{fig:training}) provides a risk-free environment for users to practice stock prediction. The area includes prediction charts with Bollinger Bands and visual cues, plus the chat box.
The bands answer the question whether prices are high or low relative to a standard 20-day window \cite{bollinger1992}. The training area allows users to explore and learn about different investment scenarios without financial consequences. 

Multiple charts support both shared-space and dispersed visual span tasks. To help novice investors build a coherent mental map, we integrated multiple charts for large datasets \cite{Drillboards}, as seen in Fig.~\autoref{fig:main}. To ensure effectiveness across task types (e.g., selection, screenshot, prediction), we primarily employed linear charts due to their simplicity and efficiency in supporting data interpretation in shared and dispersed visual contexts \cite{MultipleTimeSeries}. 
Table 1 (see Appendix) provides an overview of the functionalities and the modalities used.

%% file: sections/06-evaluation.tex
We evaluated \textit{InvestChat} with 12 novice investors in a within-subjects study.
User engagement represents a combination of both self-reported and activity-based metrics and can therefore be hard to define \cite{UserEngagement}. Thus, we combined quantitative and qualitative methods in our study. We evaluated how multimodal interaction influenced investment data exploration, measuring task completion time and using self-reported measures, such as short interviews, post-questionnaires, and the System Usability Scale (SUS) questionnaire.
We recruited 12 participants with a background in business, IT, and humanities, who were all novice investors (i.e., little to no prior investing experience). Eleven participants were in the 18---32 age range, while one was older. 
To mitigate learning-order effects, we counterbalanced the three conditions (A: Touch, B: Speech+Touch, C: Pen) via a balanced 3×3 Latin Square. Additionally, participants were always allowed to interact with the LLM assistant via typing. 

Each session started with a pre-questionnaire about demographics and prior experience with input modalities, followed by a brief introduction of \textit{InvestChat}. Then, participants had to solve three tasks per modality combination (A--C) with short interviews after each block. Following this, we asked them to solve open-ended tasks with free choice of modality (See Table 2 in Appendix). The session concluded with a final interview, a custom post-questionnaire, and the SUS questionnaire. Each session lasted circa 35 minutes.

Interestingly, although speech was reported as the hardest modality to use by eight out of 12 participants, seven also rated it as the most effective, suggesting that unfamiliarity constrained its use, but it had a strong potential. The SUS questionnaire yielded an overall score of 77, corresponding to a B+ (Excellent) rating \cite{SUScalculations}, which indicates \emph{InvestChat} was usable enough to sustain data exploration. 
We included further metrics in the appendix due to space constraints.

Overall, the results suggest multimodal interaction can facilitate user engagement by supporting precise manipulation (touch/pen) and efficient querying (natural language). Participants used modalities in complementary ways and reported that this flexibility improved the experience. In addition, our findings indicate that the dashboard supported learning and understanding of investment concepts, particularly through the integrated LLM assistant and the training area. Participants reported that the LLM was useful for explaining financial information and navigating the system itself.

The training area encouraged active learning, allowing users to predict stock trends and compare these with historical data.
Together, these features suggest that combining LLM responses with interactive training mechanisms can support novice investors in developing both conceptual understanding and analytical skills when exploring investment data. Lastly, participants reported positive feedback for the training area, with $P_4$ stating \textit{``I think the training is actually making great sense. I like that you can train your own prediction}.'' Future studies should evaluate the interaction design with broader demographics to further validate these findings.

%% file: references.bib
@String{Computer = "{IEEE} Computer" }

@String{pubACM            = {{ACM}}}

@String{pubIEEECS           = {{IEEE Computer Society}}}

@String{addrACM           = {New York, NY, USA}}

@String{addrIEEECS        = {Los Alamitos, CA, USA}}

@String{procCHI           = {Proceedings of the {ACM} Conference on Human Factors in Computing Systems}}

@ARTICLE{BeyondMouse,
  author={Lee, Bongshin and Isenberg, Petra and Riche, Nathalie Henry and Carpendale, Sheelagh},
  journal={IEEE Transactions on Visualization and Computer Graphics}, 
  title={Beyond Mouse and Keyboard: Expanding Design Considerations for Information Visualization Interactions}, 
  year={2012},
  volume={18},
  number={12},
  pages={2689--2698},
  doi={10.1109/TVCG.2012.204}
}

@inproceedings{TouchPivot,
author = {Jo, Jaemin and L'Yi, Sehi and Lee, Bongshin and Seo, Jinwook},
title = {TouchPivot: Blending WIMP \& Post-WIMP Interfaces for Data Exploration on Tablet Devices},
year = {2017},
isbn = {9781450346559},
publisher = pubACM,
address = addrACM,
url = {https://doi.org/10.1145/3025453.3025752},
doi = {10.1145/3025453.3025752},
booktitle = procCHI,
pages = {2660--2671},
numpages = {12},
location = {Denver, Colorado, USA}
}

@inproceedings{Affordances,
  author    = {Sriram Karthik Badam and Arjun Srinivasan and Niklas Elmqvist and John Stasko},
  title     = {Affordances of Input Modalities for Visual Data Exploration in Immersive Environments},
  booktitle = {Proceedings of the IEEE VIS Immersive Analytics Workshop},
  publisher = pubIEEECS,
  address   = addrIEEECS,
  year      = {2017},
  numpages  = {5},
  url={https://api.semanticscholar.org/CorpusID:20980425}
}

@ARTICLE{Talk_To_The_Wall,
  author={León, Gabriela Molina and Bezerianos, Anastasia and Gladin, Olivier and Isenberg, Petra},
  journal={IEEE Transactions on Visualization and Computer Graphics}, 
  title={Talk to the Wall: The Role of Speech Interaction in Collaborative Visual Analytics}, 
  year={2025},
  volume={31},
  number={1},
  pages={941--951},
  doi={10.1109/TVCG.2024.3456335}
  }

@inproceedings{InChorus,
author = {Srinivasan, Arjun and Lee, Bongshin and Henry Riche, Nathalie and Drucker, Steven M. and Hinckley, Ken},
title = {InChorus: Designing Consistent Multimodal Interactions for Data Visualization on Tablet Devices},
year = {2020},
isbn = {9781450367080},
publisher = pubACM,
address = addrACM,
url = {https://doi.org/10.1145/3313831.3376782},
doi = {10.1145/3313831.3376782},
booktitle = procCHI,
pages = {1–13},
numpages = {13},
}

@article{Multimodal-interaction,
title = {Multimodal interaction: Input-output modality combinations for identification tasks in augmented reality},
journal = {Applied Ergonomics},
volume = {105},
pages = {103842},
year = {2022},
issn = {0003-6870},
doi = {https://doi.org/10.1016/j.apergo.2022.103842},
url = {https://www.sciencedirect.com/science/article/pii/S000368702200165X},
author = {May Jorella Lazaro and Jaeyong Lee and Jaemin Chun and Myung Hwan Yun and Sungho Kim}
}

@article{MMI-input,
author = {Wang, Peng and Zhang, Shusheng and Bai, Xiaoliang and Billinghurst, Mark and Zhang, Li and Wang, Shuxia and Han, Dechuan and Lv, Hao and Yan, Yuxiang},
year = {2019},
month = {12},
pages = {},
title = {A gesture- and head-based multimodal interaction platform for MR remote collaboration},
volume = {105},
journal = {The International Journal of Advanced Manufacturing Technology},
doi = {10.1007/s00170-019-04434-2}
}

@article{InterChat,
author = {Chen, Juntong and Wu, Jiang and Guo, Jiajing and Mohanty, Vikram and Li, Xueming and Ono, Jorge and He, Wenbin and Ren, Liu and Liu, Dongyu},
year = {2025},
month = {05},
pages = {},
title = {InterChat: Enhancing Generative Visual Analytics using Multimodal Interactions},
journal = {Computer Graphics Forum},
volume = {44},
number = {3},
doi = {10.1111/cgf.70112}
}

@inproceedings{TimeFork,
author = {Badam, Sriram Karthik and Zhao, Jieqiong and Sen, Shivalik and Elmqvist, Niklas and Ebert, David},
title = {TimeFork: Interactive Prediction of Time Series},
year = {2016},
isbn = {9781450333627},
publisher = pubACM,
address = addrACM,
url = {https://doi.org/10.1145/2858036.2858150},
doi = {10.1145/2858036.2858150},
booktitle = procCHI,
pages = {5409--5420},
numpages = {12},
}

@ARTICLE{MultipleTimeSeries,
  author={Javed, Waqas and McDonnel, Bryan and Elmqvist, Niklas},
  journal={IEEE Transactions on Visualization and Computer Graphics}, 
  title={Graphical Perception of Multiple Time Series}, 
  year={2010},
  volume={16},
  number={6},
  pages={927--934},
  doi={10.1109/TVCG.2010.162}
  }

@inproceedings{UserEngagement,
author = {Jansen, Bernard J and Guan, Kathleen W and Salminen, Joni and Aldous, Kholoud Khalil and Jung, Soon-Gyo},
title = {What is User Engagement?: A Systematic Review of 241 Research Articles in Human-Computer Interaction and Beyond},
year = {2025},
isbn = {9798400713941},
publisher = pubACM,
address = addrACM,
url = {https://doi.org/10.1145/3706598.3713505},
doi = {10.1145/3706598.3713505},
booktitle = procCHI,
articleno = {457},
numpages = {19},
location = {
},
}

@misc{SUScalculations,
  author       = {John Bellio},
  title        = {System Usability Scale (SUS) Practical Guide for 2025},
  year         = {2024},
  month        = {December},
  day          = {19},
  howpublished = {\url{https://blog.uxtweak.com/system-usability-scale/}},
  note         = {Last accessed: 2025-06-06}
}

@misc{PeopleInvestMore,
  author       = {{Nordea Bank}},
  title        = {Young people are investing more and more --- are you following the trend?},
  year         = {2024},
  month        = {November},
  day          = {12},
  howpublished = {\url{https://www.nordea.com/en/news/young-people-are-investing-more-and-more-are-you-following-the-trend}},
  note         = {Last accessed: 2026-01-09}
}

@ARTICLE{Chowdhury_dashboard_multimodal,
  author={Chowdhury, Imran and Moeid, Abdul and Hoque, Enamul and Kabir, Muhammad Ashad and Hossain, Md. Sabir and Islam, Mohammad Mainul},
  journal={IEEE Access}, 
  title={Designing and Evaluating Multimodal Interactions for Facilitating Visual Analysis With Dashboards}, 
  year={2021},
  volume={9},
  number={},
  pages={60--71},
  doi={10.1109/ACCESS.2020.3046623},
  ISSN={2169-3536},
  month={},}

@ARTICLE{Drillboards,
  author={Shin, Sungbok and Na, Inyoup and Elmqvist, Niklas},
  journal={IEEE Transactions on Visualization and Computer Graphics}, 
  title={Drillboards: Adaptive Visualization Dashboards for Dynamic Personalization of Visualization Experiences}, 
  year={2025},
 volume={31},
  number={10},
  pages={7196--7210},
  doi={10.1109/TVCG.2025.3542606},
  ISSN={1941-0506},
  month=feb,
  }

@article{bollinger1992,
  title={Using Bollinger Bands},
  author={Bollinger, John},
  journal={Stocks \& Commodities},
  volume={10},
  number={2},
  pages={47--51},
  year={1992},
  url="https://c.mql5.com/forextsd/forum/211/Using%20Bollinger%20Bands%20by%20John%20Bollinger.pdf"
}

@article{Multimodal_Interaction_A_Review,
title = {Multimodal interaction: A review},
journal = {Pattern Recognition Letters},
volume = {36},
pages = {189-195},
year = {2014},
issn = {0167-8655},
doi = {https://doi.org/10.1016/j.patrec.2013.07.003},
url = {https://www.sciencedirect.com/science/article/pii/S0167865513002584},
author = {Matthew Turk}
}
